\documentclass[prd,showpacs,amsmath,amssymb]{revtex4}
\usepackage{graphicx,color}
\begin{document}
\title{Calculation of the Branching Ratio of $B^{-}\to h_{c}+K^{-}$ in
PQCD}

\baselineskip 22pt

%Decays of B meson into $K$ and $^{1}P_{1}$ charmonium state
%$h_{c}$  in PQCD approach}

\author{Xue-Qian Li}

\author{Xiang Liu}

\author{Yu-Ming Wang}

\affiliation{Department of physics, Nankai University, Tianjin
300071, China}
\date{\today}
\begin{abstract}
The branching ratio of $B^-\rightarrow h_c+K^-$ is re-evaluated in
the PQCD approach. In this theoretical framework all the
phenomenological parameters in the wavefunctions and Sudakov
factor are priori fixed by fitting other experimental data, and in
the whole numerical computations we do not introduce any new
parameter. Our results are consistent with the upper bounds set by
the Babar and Belle measurements.

\end{abstract}
 \maketitle

\section{introduction}

$h_{c}$ of $^{1}P_{1}$ is the spin-partner of $\eta_c$ in the
charmomium family, however, it seems to $h_c$ behaves very
differently from other family members. Searching for $h_c$, as
well as $h_b$ is a challenging task for both experimentalists and
theorists of high energy physics \cite{barnes}. $h_{c}$ was first
seen at the CERN Intersecting Storage Rings (ISR) \cite{cern-isr},
and later the  E760 group at Fermilab reported observation of
$h_{c}$ by studying the $p\bar{p}\to h_{c}\to \pi^{0}J/\psi$
\cite{fermilab-E760}. However, the E835 group claimed that they
did not see $h_{c}$ in this channel. Instead, E835 reported
observation of $h_{c}$ in another channel $p\bar{p}\to h_{c}\to
\gamma\eta_{c}\to \gamma\gamma\gamma$ \cite{fermilab-E835}.
Recently, the Babar collaboration set an upper bound for the
production rate of $h_c$ as $3.4\times 10^{-6}$ for $BR(B^{-}\to
h_{c} K^{-})\times BR(h_{c} \to J/\psi \pi^{+}\pi^{-})$ at $90\%$
C.L. \cite{Babar-hc}. By contrast, the Belle collaboration
reported null result on searching for $h_c$ via $B^{\pm}\to h_c
K^{\pm}$ and gave an upper bound on the branching ratio as
$BR(B^{+}\to h_c K^{+})<3.8\times 10^{-5}$ \cite{Belle-hc}. The
CLEO Collaboration has announced observation of $h_{c}$ in the
decay $\psi(2S)\to \pi^{0}h_{c}\to\pi^{0}\gamma\eta_{c}$ with a
very small branching ratio \cite{hc-CLEO-1,hc-CLEO-2}. As a
comparison, so far, the BES II collaboration has not seen $h_c$
yet, and searching for it may be an important issue for the
upgraded stage BES III.

One needs to understand the experimental status about the
production rate of $h_c$.  Moreover, a relatively larger
production rate may be associated to new physics
\cite{Demin,Aydin,Bi}, therefore study on $h_c$ production may
provide a chance to investigate effects of new physics, such as
supersymmetry at lower energy scales. Of course, before invoking
for new physics, one needs to more accurately estimate the
production rate of $h_c$ in the standard model (SM).

Suzuki suggested to look for $h_{c}$ at $B^{+}\to h_{c}K^{+}\to
\gamma\eta_{c}K^{+}\to \gamma(K\bar{K}\pi)K^{+}$ and if
approximately  $BR(B^{+}\to h_{c}K^{+})\approx BR(B^{+}\to
\chi_{c0}K^{+})$, he estimated the cascade branching ratio as
$2\times 10^{-5}$ \cite{suzuki}. Gu considered another decay chain
and estimated the branching ratio  $BR(B^{+}\to h_{c}K^{+}\to
\gamma\eta_{c}K^{+}\to \gamma(K_{S}^{0}K^{+}\pi^{-}+c.c.)K^{+}
\to\gamma(\pi^{+}\pi^{-}K^{+}\pi^{-}+c.c.)K^{+})$ as
$3.5\times10^{-6}$\cite{gu}.

In the decay of B-meson into charmonium is the so-called internal
emission process where due to the color matching the process is
suppressed compared to the external emission. Moreover, the
non-factorizable effects would further change the contribution of
the internal process besides a color factor of 1/3 \cite{Cheng}.
Therefore it seems that the smallness of $B\rightarrow h_c+K$ is
natural. However, one could not reproduce experimental value of
the $B^{-}\to \chi_{c0}K^{-} $ decay rate \cite{colangelo 1} in
the QCD improved factorization, so that the authors of
\cite{colangelo 1} suggested to consider re-scattering effects in
B meson decay. They claimed that \cite{colangelo 1} a larger
branching ratio for $B^{-}\to \chi_{c0} K^{-}$ which is consistent
with data, was obtained. Motivated by the same idea, they applied
the same scenario to investigate the re-scattering effects for
$B^{-}\to h_c K^{-}$ case, i.e. supposing that the decay $B^{-}\to
h_{c}K^{-}$ occurs via the the subsequent re-scattering effect of
$D_{s}^{(*)}-D^{(*)0}$, the products of decay $B^{-}\to
D_{s}^{(*)}D^{(*)0}$ \cite{colangelo 2}. They obtained a branching
ratio ($BR(B^{-}\to h_c K^{-})=(2\sim10) \times 10^{-4} $) which
is much larger than the upper bound set by the Belle
collaboration. It is very possible that the branching ratio of
$B^{-}\to h_c K^{-}$ was overestimated in their work due to
uncontrollable theoretical uncertainties, such as some input
parameters and basic assumptions adopted in their calculation,
which were comprehensively discussed in their paper.

Even though we may suppose that we have full knowledge on the weak
and strong interactions at the quark level and can derive the
quark-transition amplitude, the most difficult part is evaluation
of the hadronic matrix elements of the exclusive processes. In
fact, at present, a complete calculation of $B^{\pm}\to h_c
K^{\pm}$ based on an underlying theoretical framework is absent.
The perturbative QCD (PQCD) approach is believed to be successful
for estimating transition rates of B and D into light mesons
\cite{pqcd classic}, even though there is still dispute about its
applicability \cite{Sachrajda}. The authors of ref.
\cite{hc-amplitude} applied the PQCD approach to study $B^{-}\to
\chi_{c0} K^{-}$ and also obtained results which satisfactory
comply with data. Therefore we have reason to believe that it is
appropriate to analyze $B^{\pm}\to h_c K^{\pm}$ in this framework.
In this work, we calculate decay rate of $B^{-}\to h_{c}K^{-}$ in
the PQCD.

Generally, for two-body non-leptonic decays of B meson, both
factorizable and non-factorizable diagrams contribute to the
transition amplitudes, however, for  $B^{-}\to h_{c}K^{-}$, the
contributions from factorizable diagrams disappear since the
conservation of G parity leads to
$\langle0|\bar{c}\gamma_{\mu}\gamma_5 c|h_c\rangle=0$ \cite{K.C.
Yang}. Therefore, the decay rate of $B^{\pm}\to h_c K^{\pm}$ is
much different from that in $B^{\pm}\to J/\psi K^{\pm}$
\cite{psi,hc-amplitude} where  both factorizable and
non-factorizable diagrams contribute.

To be more precise than the qualitative understanding, one needs
to calculate the non-factorizable contribution where the QCD
effects are accounted. Moreover, since the process is not
factorizable, the convolution integral would involve the initial
B-meson and all the two produced mesons altogether.

Our numerical results indeed indicate that the order of magnitude
of $B^{-}\to h_c K^{-}$ should be of order of  $3.6\times
10^{-5}$, which is smaller than the upper bound set by the Babar
collaboration \cite{Babar-hc} and slightly below the upper bound
given by the Belle collaboration \cite{Belle-hc}. It is noticed
that our result about $B(B^{-}\to h_c K^{-})$ is smaller than that
estimated by the authors of \cite{suzuki,gu,colangelo 1} and more
consistent with data.

Indeed, more decisive conclusion should be made as more accurate
data are accumulated by BES III, CLEO,  Babar, Belle and even the
LHCb.

The structure of this paper is organized as follows. After this
introduction, we formulate the decay amplitude of $B^{-}\to
h_{c}K^{-}$ in the PQCD approach. Then  we present our numerical
results along with all the input parameters in Sect. III. The last
section is devoted to our conclusion and discussion. Some tedious
expressions are collected in the Appendix.

\section{Formulation}

\begin{figure}[htb]
\begin{center}
\begin{tabular}{cccccccc}
\scalebox{0.8}{\includegraphics{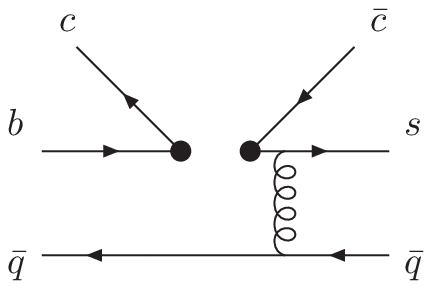}}&&&&&&
\scalebox{0.8}{\includegraphics{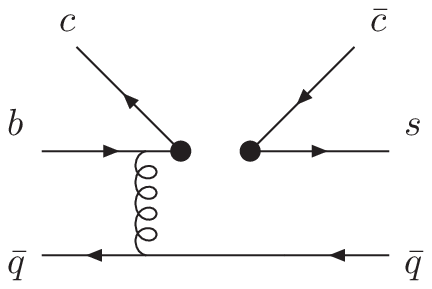}}\\
(a)&&&&&&(b)
\\\scalebox{0.8}{\includegraphics{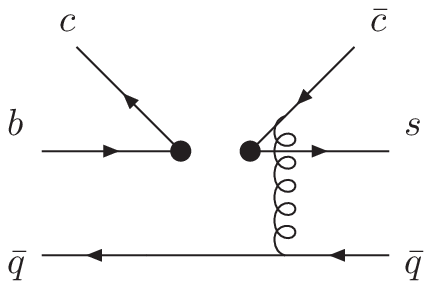}}&&&&&&
\scalebox{0.8}{\includegraphics{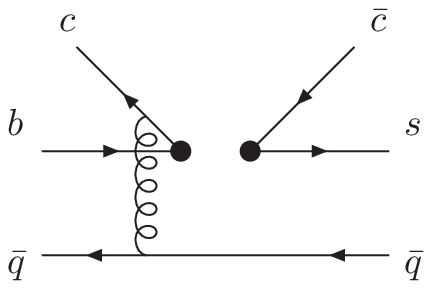}}\\
(c)&&&&&&(d)\\
\end{tabular}
\end{center}
\caption{Feynman diagrams correspond to the calculation of hard
amplitudes in $B\to h_{c}K$.}\label{hard}
\end{figure}

The quark-diagrams which contribute to the transition amplitude of
$B^{-}\to (^{1}P_{1})h_{c}+K^{-}$ are displayed in Fig.
\ref{hard}. As has been discussed in the introduction, the
factorized diagrams {Fig. \ref{hard} (a),(b)} do not contribute to
the amplitude because the the G parities of axial meson $h_{c}$
and axial current are mismatched and in the flavor SU(3) symmetry
limit, the corresponding hadronic matrix element is
forbidden\cite{colangelo 1, K.C. Yang}.

The effective Hamiltonian relevant to $B^{-}\to h_{c}K^{-}$ decay
in the SM is written as \cite{effective-H}
\begin{eqnarray}
\mathcal{H}_{eff}=\frac{G_{F}}{\sqrt{2}}\bigg[ V_{cb}V_{cs}^{*}
\big(\mathcal{C}_{1}(\mu)\mathcal{O}_{1}+\mathcal{C}_{2}(\mu)\mathcal{O}_{2}\big)
-V_{tb}V_{ts}^{*}\sum^{10}_{i=3}\mathcal{C}_{i}(\mu)\mathcal{O}_{i}\bigg],
\end{eqnarray}
where $C_{i}(\mu)$ are the Wilson coefficients and
$\mathcal{O}_{i}$ are the relevant operators defined as
\begin{eqnarray*}
\mathcal{O}_{1}&=&(\bar{s}_{\alpha}c_{\beta})_{V-A}(\bar{c}_{\beta}b_{\alpha})_{V-A},\;\;\;
\mathcal{O}_{2}=(\bar{s}_{\alpha}c_{\alpha})_{V-A}(\bar{c}_{\beta}b_{\beta})_{V-A},\nonumber\\
\mathcal{O}_{3(5)}&=&(\bar{s}_{\alpha}b_{\alpha})_{V-A}\sum_{q}(\bar{q}_{\beta}q_{\beta})_{V-A(V+A)},\;\;\;
\mathcal{O}_{4(6)}=(\bar{s}_{\alpha}b_{\beta})_{V-A}\sum_{q}(\bar{q}_{\beta}q_{\alpha})_{V-A(V+A)},\\
\mathcal{O}_{7(9)}&=&\frac{3}{2}(\bar{s}_{\alpha}b_{\alpha})_{V-A}\sum_{q}e_{q}(\bar{q}_{\beta}q_{\beta})_{V+A(V-A)},\;\;\;
\mathcal{O}_{8(10)}=\frac{3}{2}(\bar{s}_{\alpha}b_{\beta})_{V-A}\sum_{q}e_{q}(\bar{q}_{\beta}q_{\alpha})_{V+A(V-A)},\nonumber
\end{eqnarray*}
with $\alpha$, $\beta$ being the color indices. The explicit
expressions of the Wilson coefficients appearing in the above
equations can be found in Ref. \cite{effective-H}.

We define, in the rest frame of the B meson, $p$, $p'$ and $q$ to
be the four-moneta of $B$, $K$ and $h_{c}$, $k_{1(2)}$,
$k_{1(2)}'$ and $q_{1(2)}$ to be the momenta of the valence quarks
inside B($b(\bar{q})$), $K$ ($s(\bar{q})$ ), and  $h_{c}$
($c(\bar{c})$ ), respectively. Then we parameterize the light cone
momenta with all the light quarks and mesons being treated  as
massless
\begin{eqnarray*}
p&=&\frac{m_{B}}{\sqrt{2}}(1,1,\mathbf{0}_{\mathrm{T}})=(p^{+},p^{-},\mathbf{0}_{\mathrm{T}}),\qquad
p'=\frac{m_{B}}{\sqrt{2}}(0,1-r^{2},\mathbf{0}_{\mathrm{T}}),\qquad
q=\frac{m_{B}}{\sqrt{2}}(1,r^{2},\mathbf{0}_{\mathrm{T}})=(q^{+},q^{-},\mathbf{0}_{\mathrm{T}}),\\
k_{1}&=&{(x_{1}p^{+},p^{-},\mathbf{k}_{1\mathrm{T}})},\qquad
k_{2}={(\bar{x}_{1}p^{+},0,-\mathbf{k}_{1\mathrm{T}})},\qquad
k_{1}'=(x_{1}'p'^{+},x_{1}'p'^{-},\mathbf{k}_{1\mathrm{T}}'),\\
k_{2}'&=&(\bar{x_{1}'}p'^{+},
\bar{x_{1}'}p'^{-},-\mathbf{k}'_{1\mathrm{T}}),\qquad
q_{1}=(yq^{+},yq^{-},\mathbf{q}_{T}),\qquad
q_{2}=(\bar{y}q^{+},\bar{y}q^{-}, -\mathbf{q}_{\mathrm{T}}),
\end{eqnarray*}
where the mass ratio $r$ is set as $r=m_{h_{c}}/m_{B}$. $x_{i}$
and $x_{i}'$ are the fractions of the longitudinal momenta of the
valence quarks. The superscripts "$+$" (plus) and "$-$" (minus)
mean that the three-momentum is parallel or anti-parallel to the
positive $z$ direction which is defined as the direction of the
three-momentum of the produced $h_c$. $\mathbf{k}_{1\mathrm{T}}$,
$\mathbf{k}_{1\mathrm{T}}'$ and $\mathbf{q}_{\mathrm{T}}$ are the
transverse momenta of the valence quarks inside B, $K$ and $h_{c}$
respectively.

Therefore, we concentrate on the calculations of the
non-factorizable diagrams {Fig. \ref{hard} (c),(d).} Then we
divide the operators appearing in the effective Hamiltonian into
two categories according to their chirality, i.e. $(V-A)\otimes
(V-A)$ and $(V-A)\otimes (V+A)$.

For the type of $(V-A)\otimes (V\mp A)$, the hard kernels of Fig.
\ref{hard} (c) are respectively written as
\begin{eqnarray}
H^{(c,1)}_{\alpha\beta\rho\alpha'\beta'\rho'}(p,p',q)&=&\bigg[ig_{s}\gamma_{\nu}\frac{i}{\not
q_{1}-\not k _{2}+\not
k'_{2}-m_{c}}\gamma_{\mu}(1-\gamma_{5})\bigg]_{\rho\alpha}[\gamma^{\mu}(1-\gamma_{5})]
_{\alpha'\rho'}[ig_{s}\gamma^{\nu}]_{\beta\beta'}\frac{-i}{(k_{2}-k_{2}')^2},\\
H^{(c,2)}_{\alpha\beta\rho\alpha'\beta'\rho'}(p,p',q)&=&-2\bigg[ig_{s}\gamma_{\nu}\frac{i}{\not
q_{1}-\not k _{2}+\not
k'_{2}-m_{c}}(1-\gamma_{5})\bigg]_{\rho\alpha}[(1+\gamma_{5})]
_{\alpha'\rho'}[ig_{s}\gamma^{\nu}]_{\beta\beta'}\frac{-i}{(k_{2}-k_{2}')^2}\label{hard-c},
\end{eqnarray}
where the factor '-2' in eq. (\ref{hard-c}) comes from the Fierz
transformation on the $(V-A)\otimes (V+A)$ operators.

Similarly, for Fig. \ref{hard} (d), we have
\begin{eqnarray}
H^{(d,1)}_{\alpha\beta\rho\alpha'\beta'\rho'}(p,p',q)&=&[\gamma_{\mu}(1-\gamma_{5})]
_{\rho\alpha}\bigg[ig_{s}\gamma^{\mu}(1-\gamma_{5})\frac{i}{-(\not
q_{2}-\not k _{2}+\not
k'_{2})-m_{c}}\gamma_{\nu}\bigg]_{\alpha'\rho'}[ig_{s}\gamma^{\nu}]_{\beta\beta'}
\frac{-i}{(k_{2}-k_{2}')^2},\\
H^{(d,2)}_{\alpha\beta\rho\alpha'\beta'\rho'}(p,p',q)&=&-2[\gamma_{\mu}(1-\gamma_{5})]
_{\rho\alpha}\bigg[ig_{s}\gamma^{\mu}(1+\gamma_{5})\frac{i}{-(\not
q_{2}-\not k _{2}+\not
k'_{2})-m_{c}}\gamma_{\nu}\bigg]_{\alpha'\rho'}[ig_{s}\gamma^{\nu}]_{\beta\beta'}
\frac{-i}{(k_{2}-k_{2}')^2}.
\end{eqnarray}
In the above expressions, indices 1,2 denote the contributions
from $(V-A)\otimes (V-A)$ and $(V-A)\otimes (V+A)$ operators
respectively.

Finally, for each type of operators we obtain the decay amplitude
$M_{i}^{(j)}$ which is a  sum of the two non-factorizable diagrams.
For saving the space, we collect them in Appendix A. The wave
functions relevant to calculation is given in Appendix B.

The decay width of $B^{-}\to h_{c}K^{-}$ is written as
\begin{eqnarray}
\Gamma=\frac{G_{F}^2}{2}\frac{|\mathbf{p}_{f}|}{8\pi
m_{B}^{2}}|\mathcal{M}|^2
\end{eqnarray}
with the total amplitude is the sum of $M_{1,2}(c,d)$ as
$$\mathcal{M}=M_{1}^{(c)}+M_{2}^{(c)}+M_{1}^{(d)}+M_{2}^{(d)}.$$
$\mathbf{p}_{f}$ denotes the three-momentum of the produced meson in
the center-of-mass frame of B meson. All the explicit expressions of
$M_i$ are collected in Appendix A in order to shorten the text and
focus on the physics contents.

\section{Numerical results}

The input parameters used in the text are given below: $m_{B}=5.279$
GeV, $m_{K}=0.494$ GeV \cite{PDG}. $m_{h_{c}}=3.524$ GeV
\cite{hc-CLEO-1,hc-CLEO-2}. For the CKM mixing parameters, we take
$s_{12} = 0.2243$, {$s_{23} = 0.0413$,} $s_{13} = 0.0037$ and
$\delta_{13} = 1.05$ \cite{PDG}. The parameters appeared in the wave
functions are put in the Appendix B following the corresponding
wavefunctions.

Using the above parameters, finally one obtains the branching
ratio of $B^{-}\to h_{c}K^{-}$ (in the numerical calculation, we
take the central values listed in the data book for the input
parameters)
$$BR(B^{-}\to h_{c}K^{-})=3.6\times 10^{-5}.$$

Since all the parameters adopted in our numerical computations are
included in the wavefunctions of the concerned hadrons and the
Sudakov factor which are obtained by fitting data of some well
measured processes, thus the error of our final result is due to
the uncertainties in such fitting. Because of the uncontrollable
factors, one cannot expect to get very accurate value at the
present stage. Therefore, we only keep two significant figures
without explicitly marking out the error range. But in the last
section for discussion, we will present a rough error estimate
which may make sense.

There are some theoretical uncertainties in our calculations. One of
them comes from the next to leading order corrections to the hard
amplitudes\cite{Y.L Shen}. In view of this point, we check the
sensitivity of the decay rate with different choices of hard scales,
i.e. we set the hard scales as
\begin{eqnarray}
&&max(0.75\sqrt{A_c},0.75\sqrt{B_c},1/b_1,1/b'_1,1/b_q,1/|{\bf{b_1-b_q}}|)\leq
t_c \leq
max(1.25\sqrt{A_c},1.25\sqrt{B_c},1/b_1,1/b'_1,1/b_q,1/|{\bf{b_1-b_q}}|)\nonumber \\
&&max(0.75\sqrt{A_d},0.75\sqrt{B_d},1/b_1,1/b'_1,1/b_q,1/|{\bf{b_1+b_q}}|)\leq
t_d \leq
max(1.25\sqrt{A_d},1.25\sqrt{B_d},1/b_1,1/b'_1,1/b_q,1/|{\bf{b_1+b_q}}|),
\nonumber
\end{eqnarray}
and other parameters are fixed. Then we can obtain the branching
ratio of $B^{-}\to h_{c}K^{-}$ and the error may be  a few percents
and the whole result is not sensitive to the change of hard scales.

Another uncertainty comes from the non-perturbative parameters in
meson wavefunctions, such as $\omega_b$ in  $B$ meson wavefunctions,
although they are determined directly from previous experiments or
some non-perturbative methods like QCD sum rules. Here we vary the
value of $\omega_b$ at the range $0.32\sim 0.48$ GeV, with other
parameters fixed, as Ref. \cite{Y.L Shen} did, then we find that the
errors can be as large as 20\%. It can be observed that the decay
rate is relatively more dependent on the value of $\omega_b$.

In Ref. \cite{suzuki}, the author proposed that $h_{c}\to
\eta_{c}\gamma$ is a promising mode to look for $h_c$ and presented
a theoretical estimation on the branching ratio as
$BR(h_{c}\to\eta_{c}\gamma)=0.50\pm 0.11$. Combining this branching
ratio with $BR(\eta_{c}\to K\bar{K}\pi)=(5.7\pm 1.6)\%$, one could
predict the branching ratio of the decay chain $B^{-}\to
h_{c}K^{-}\to (\eta_{c}\gamma) K^{-}\to ((K\bar{K}\pi)\gamma)K^{-}$
as
\begin{eqnarray}
BR(B^{-}\to h_{c}K^{-}\to (\eta_{c}\gamma) K^{-}\to
((K\bar{K}\pi)\gamma)K^{-})=1.0\times10^{-6}.
\end{eqnarray}

\section{Discussion and conclusion}
In this work, we employ the framework of PQCD approach to calculate
the transition rate of $B^{-}\to h_{c}K^{-}$ and predict the
branching ratio to be about $3.6 \times 10^{-5}$.

{It is observed that only non-factorizable diagrams contribute to
the decay amplitude, because this transition is a G-parity violation
process. So one can expect it to be relatively suppressed compared
with $B^{-}\to J/\psi K^{-}$ which is about $1.00 \times 10^{-3}$
\cite{PDG}.}  In our work, we estimate the ratio and obtain results
which are reasonably consistent with data. In this way, we not only
naturally reproduce the production rate of $h_c$ at Babar and Belle,
but also have a chance to study the non-factorizable diagrams.
Usually, in most processes, both factorizable and non-factorizable
diagrams contribute and it is hard to separate their contributions.
Thus when comparing with data, there is an obvious uncertainty.
However, this case is an optimistic place where only
non-factorizable diagrams contribute. It enables us to uniquely
investigate the non-factorizable contributions to the weak
transitions and testify applicability of the theory, PQCD.

Our prediction is consistent with the upper bound reported by the
Belle and Babar collaborations \cite{Belle-hc} at the level of
order of magnitude. Thus we can conclude that the PQCD is
applicable to the process, especially to deal with the
non-factorizable diagrams.

In fact, all these phenomenological parameters in the wavefunctions
of $B,K,h_c$ are priori determined and the choice of the
factorization scale follows the conventional way. In our numerical
computations, we do not introduce any free parameter to adjust.
Therefore the error of the final result obtained in this work is due
to the uncertainties included in the wavefunctions of the concerned
hadrons and the Sudakov factors which are expressed in terms of a
few phenomenological parameters.

As the input parameters which exist in the wavefunctions of the
concerned hadrons, vary within a reasonable range, an error of
about 20\% is expected. On other aspect, it would be difficult to
make a precise estimate on the uncertainty because it is
determined by non-perturbative QCD effects, thus only the order of
magnitude of the results can be trusted. Even then, our
theoretical predictions are more consistent with the present data,
i.e. the upper bounds set by the experiments. It seems that based
on the PQCD approach, the obtained results are more consistent
with data than the earlier analyses \cite{suzuki,gu,colangelo 2},
thus one expects that the theoretical framework may reflect the
real physics picture, at least works in this case.

If the future experiments further reduce the upper bound and the
theoretical estimate on the branching ratio of $B^-\rightarrow
h_c+K^-$ cannot tolerate the new data even considering the range
for the input parameters, one would confront a serious challenge
to our understanding of the physics mechanisms, therefore, we hope
that by improvement to reduce both statistical and systematic
errors, new measurements of the Babar and Belle collaborations
will provide important information towards the answer, otherwise
we may need to wait for the data of the future LHC-b.

To be more accurate in both theory and experiment, we need to wait
for more data and improvement in the method. Indeed, a favorable
channel for observing $h_c$ is decay of $\psi(2S)$. However,
unfortunately, the BES collaboration has not observed this mode and
one needs to wait for the upgraded stage, i.e BES III which would
accumulate much larger database in future and then we can further
investigate the consistency between theory and experiments.

\section*{Acknowledgements}
This work is  supported by the National Natural Science Foundation
of China(NNSFC). The authors would like to thank C.D. L\"{u}, W.
Wang and T. Li for helpful discussions. {We also would like to thank
S.F. Tuan for useful comments.}

\section*{Appendix A: Some relevant functions appeared in the text}

The explicit expressions of decay amplitude $M_{i}^{(j)}$ appeared
in the text can be written as

\begin{eqnarray}
M_{1}^{(c)}&=& \frac{8} {(2N_{c})^{3/2}}\int d x_{1}\int d
x_{1}'\int dy \int b_{1}db_{1}\int b_{q}db_{q}\int d\theta
[V_{cb}V_{cs}^{*}\mathcal{C}_{2}(\mu)-V_{tb}V^{*}_{ts}(\mathcal{C}_{4}(\mu)+\frac{3}{2}e_{c}\mathcal{C}_{10}(\mu))]
\phi_{B}(x_{1},b_{1})\phi_{h_c}^{
\parallel}(y)\nonumber\\&&\times\bigg[\phi_{K}^{A}(x_{1}')\mathbb{KA}^{(c)}_{1}+\phi_{K}^{P}(x_{1}')\mathbb{KP}^{(c)}_{1}
+\phi_{K}^{T}(x_{1}')\mathbb{KT}^{(c)}_{1}\bigg]\alpha_{s}(t_{c})\exp[-S(t_{c})]\Omega_{c}(x_{1},x_{1}'
,y,b_{1},b_{q}),
\end{eqnarray}

\begin{eqnarray}
M_{2}^{(c)}&=& \frac{-16} {(2N_{c})^{3/2}}\int d x_{1}\int d
x_{1}'\int dy \int b_{1}db_{1}\int b_{q}db_{q}\int d\theta
[\mathcal{C}_{6}(\mu)+\frac{3}{2}e_{c}\mathcal{C}_{8}(\mu)]{(-V_{tb}V^{*}_{ts})}
\phi_{B}(x_{1},b_{1})\phi_{h_c}^{
\parallel}(y)\nonumber\\&&\times\bigg[\phi_{K}^{A}(x_{1}')\mathbb{KA}^{(c)}_{2}+\phi_{K}^{P}(x_{1}')\mathbb{KP}^{(c)}_{2}
+\phi_{K}^{T}(x_{1}')\mathbb{KT}^{(c)}_{2}\bigg]\alpha_{s}(t_{c})\exp[-S(t_{c})]\Omega_{c}(x_{1},x_{1}'
,y,b_{1},b_{q}),
\end{eqnarray}

\begin{eqnarray}
M_{1}^{(d)}&=& \frac{8} {(2N_{c})^{3/2}}\int d x_{1}\int d
x_{1}'\int dy \int b_{1}db_{1}\int b_{q}db_{q}\int d\theta
[V_{cb}V_{cs}^{*}\mathcal{C}_{2}(\mu)-V_{tb}V^{*}_{ts}(\mathcal{C}_{4}(\mu)+\frac{3}{2}e_{c}\mathcal{C}_{10}(\mu))]
\phi_{B}(x_{1},b_{1})\phi_{h_c}^{
\parallel}(y)\nonumber\\&&\times\bigg[\phi_{K}^{A}(x_{1}')\mathbb{KA}^{(d)}_{1}+\phi_{K}^{P}(x_{1}')\mathbb{KP}^{(d)}_{1}
+\phi_{K}^{T}(x_{1}')\mathbb{KT}^{(d)}_{1}\bigg]\alpha_{s}(t_{d})\exp[-S(t_{d})]\Omega_{d}(x_{1},x_{1}'
,y,b_{1},b_{q}),
\end{eqnarray}

\begin{eqnarray}
M_{2}^{(d)}&=& \frac{-16} {(2N_{c})^{3/2}}\int d x_{1}\int d
x_{1}'\int dy \int b_{1}db_{1}\int b_{q}db_{q}\int d\theta
[\mathcal{C}_{6}(\mu)+\frac{3}{2}e_{c}\mathcal{C}_{8}(\mu)]{(-V_{tb}V^{*}_{ts})}
\phi_{B}(x_{1},b_{1})\phi_{h_c}^{
\parallel}(y)\nonumber\\&&\times\bigg[\phi_{K}^{A}(x_{1}')\mathbb{KA}^{(d)}_{2}+\phi_{K}^{P}(x_{1}')\mathbb{KP}^{(d)}_{2}
+\phi_{K}^{T}(x_{1}')\mathbb{KT}^{(d)}_{2}\bigg]\alpha_{s}(t_{d})\exp[-S(t_{d})]\Omega_{d}(x_{1},x_{1}'
,y,b_{1},b_{q}),
\end{eqnarray}
where the explicit expressions of $\mathbb{KA}^{(j)}_{i}$,
$\mathbb{KP}^{(j)}_{i}$ and $\mathbb{KT}^{(j)}_{i}$ which come
from the contraction of hard kernel and hadronic wave functions
are given below. Here the  meaning of indices i, j have been shown
in the text.
\begin{eqnarray}
\mathbb{KA}_{1}^{(c)}&=&{-16 mb^4 r^2 (-1+r^2)(-x_1+x_1'-y)},\\
\mathbb{KP}_{1}^{(c)}&=&{-16 mb^3 m_{0}^{K} r^2 (-x_1+x_1'-y)},\\
\mathbb{KT}_{1}^{(c)}&=&{16 mb^3 m_{0}^{K} r^2 (-x_1+x_1'-y)},
\\
\mathbb{KA}_{2}^{(c)}&=&\mathbb{KA}_{1}^{(d)}=0,\\
\mathbb{KP}_{2}^{(c)}&=&{-8 mb^3 m_{0}^{K} r^2 (-x_1+x_1'-y)},\\
\mathbb{KT}_{2}^{(c)}&=&{-8 mb^3 m_{0}^{K} r^2 (-x_1+x_1'-y)},
\\
\mathbb{KP}_{1}^{(d)}&=&{16 mb^3 m_{0}^{K} r^2 (-1-x_1+x_1'+y)}
,\\
\mathbb{KT}_{1}^{(d)}&=&{16 mb^3 m_{0}^{K} r^2 (-1-x_1+x_1'+y)}
,\\
\mathbb{KA}_{2}^{(d)}&=&{8 mb^4 r^2 (-1+r^2)(-1-x_1+x_1'+y)}
,\\
\mathbb{KP}_{2}^{(d)}&=&{8 mb^3 m_{0}^{K} (-1-x_1+x_1'+y)}
,\\
\mathbb{KT}_{2}^{(d)}&=&{-8 mb^3 m_{0}^{K} (-1-x_1+x_1'+y)}.
\end{eqnarray}

The explicit forms of $\Omega_{c,d}(x_{1},x_{1}',y,b_{1},b_{q})$
which come from Fourier transformation to products of propagators
corresponding to quark and gluon are listed as follow
\begin{eqnarray}
\Omega_{c}(x_{1},x_{1}'
,y,b_{1},b_{q})&=&\Big\{K_{0}(\sqrt{\mathcal{A}_{c}}|\mathbf{b}_{q}|)\theta(\mathcal{A}_{c})
+\frac{\pi}{2}[-N_{0}(\sqrt{\mathcal{A}_{c}}|\mathbf{b}_{q}|)+i
J_{0}(\sqrt{\mathcal{A}_{c}}|\mathbf{b}_{q}|)]\theta(-\mathcal{A}_{c})\Big\}
\nonumber\\&&\times K_{0}(\sqrt{\mathcal{B}_{c}}|\mathbf{b}_{1}-\mathbf{b}_{q}|),\\
\Omega_{c}(x_{1},x_{1}' ,y,b_{1},b_{q})&=&
\Big\{K_{0}(\sqrt{\mathcal{A}_{d}}|\mathbf{b}_{q}|)\theta(\mathcal{A}_{d})
+\frac{\pi}{2}[-N_{0}(\sqrt{\mathcal{A}_{d}}|\mathbf{b}_{q}|)+i
J_{0}(\sqrt{\mathcal{A}_{d}}|\mathbf{b}_{q}|)]\theta(-\mathcal{A}_{d})\Big\}
\nonumber\\&&\times
K_{0}(\sqrt{\mathcal{B}_{d}}|\mathbf{b}_{1}+\mathbf{b}_{q}|),
\end{eqnarray}
with
\begin{eqnarray}
&&{A_c=m_c^2-(x_1+y-1)((1-x'_1)(1-r^2)+y r^2)m_b^2}, \qquad
B_c=(1-x_1)(1-x'_1)(1-r^2)m_b^2, \nonumber \\
&&{A_d=m_c^2-(x_1-y)((1-x'_1)(1-r^2)+(1-y) r^2)m_b^2}, \qquad
B_d=(1-x_1)(1-x'_1)(1-r^2)m_b^2. \nonumber
\end{eqnarray}
Here $J_{i}$, $N_{i}$ are 'ith' order Bessel functions of first
and second kind respectively, $K_{i}$ denotes 'ith' order modified
Bessel functions.

The explicit expressions of Sudakov factor coming from the
resummation of double logarithm appeared in high order radiative
corrections to the diagrams are also given below
\begin{eqnarray}
S(t_{c})&=&s\big({(1-x_{1})p^{+}},b_{1}\big)+{s\big({x_{1}'
p'^{-}},b_{1}'\big)}
+s\big({(1-x_{1}')p'^{-}},b_{1}'\big)\nonumber\\&&-\frac{1}{\beta_{1}}\bigg[\ln\frac{-\ln(t_{c}/\Lambda_{QCD})}{-\ln(b_{1}\Lambda_{QCD})}+
\ln\frac{-\ln(t_{c}/\Lambda_{QCD})}{\ln(b_{1}'\Lambda_{QCD})}+\ln\frac{-\ln(t_{c}/\Lambda_{QCD})}{\ln(b_{q}\Lambda_{QCD})}
\bigg],
\\
S(t_{d})&=&s\big({(1-x_{1})p^{+}},b_{1}\big)+{s\big({x_{1}'
p'^{-}},b_{1}'\big)}
+s\big({(1-x_{1}')p'^{-}},b_{1}'\big)\nonumber\\&&-\frac{1}{\beta_{1}}\bigg[\ln\frac{-\ln(t_{d}/\Lambda_{QCD})}{-\ln(b_{1}\Lambda_{QCD})}+
\ln\frac{-\ln(t_{d}/\Lambda_{QCD})}{\ln(b_{1}'\Lambda_{QCD})}+\ln\frac{-\ln(t_{d}/\Lambda_{QCD})}{\ln(b_{q}\Lambda_{QCD})}
\bigg],
\end{eqnarray}
where
\begin{eqnarray}
t_c=max(\sqrt{A_c},\sqrt{B_c},1/b_1,1/b'_1,1/b_q,1/|{\bf{b_1-b_q}}|),
\qquad
t_d=max(\sqrt{A_d},\sqrt{B_d},1/b_1,1/b'_1,1/b_q,1/|{\bf{b_1+b_q}}|).
\nonumber
\end{eqnarray}
 The explicit expressions for the Sudakov factors are given
in Ref.\cite{Lip,Li-sterman} as
\begin{eqnarray}
&&s(\omega,Q)=\int^Q_\omega{dp\over p}\bigg[\ln\big({Q\over
p}\big)A[\alpha_s(p)]+B[\alpha_s(p)]\bigg],\nonumber \\
&&A=C_F{\alpha_s\over \pi}+\bigg[{67\over 9}-{\pi^2\over 3}-{10\over
27}n_f+{8\over 3}\beta_0ln\big({e^{\gamma_E}\over
2}\big)\bigg]({\alpha_s\over
\pi})^2,\nonumber \\
&&B={2\over 3}{\alpha_s\over \pi}\ln\big({e^{2\gamma_E-1}\over
2}\big),\nonumber \\
&&\gamma_q (\alpha_s(\mu)) = -\alpha_s(\mu)/\pi,\nonumber\\
&&\beta_0={33-2n_f\over 12}, \nonumber
\end{eqnarray}
where $\gamma_E$ is the Euler constant. $n_f$  is the flavor
number, and $\gamma_q$ is the anomalous dimension. We will take
$n_f$ equal to 4 in our numerical calculations.

\subsection*{Appendix B: Wave functions relevant to calculation}

The B meson light cone wavefunction is usually written as
\cite{B-matrix1}
\begin{eqnarray}
\int^{1}_{0}\frac{d^{4}z}{(2\pi)^4}e^{i{{k}}\cdot{{z}}} \langle
0|\bar{q}_{\beta}(0)b_{\alpha}({z})|{B}(p)\rangle=-\frac{i}{\sqrt{2N_{c}}}\bigg\{(\not
p +m_{B})\gamma_{5}\Big[\phi_{B}(\mathbf{k})-\frac{\not n -\not
v}{\sqrt{2}}\bar{\phi}_{B}(\mathbf{k})\Big]\bigg\}_{\alpha\beta},\label{b-wave}
\end{eqnarray}
where $n\equiv (1,0,\mathbf{0}_{\mathrm{T}})$ and $v\equiv
(0,1,\mathbf{0}_{\mathrm{T}})$ denote the unit vectors
corresponding to the "plus" and "minus" directions respectively.
In eq. (\ref{b-wave}), two different Lorentz structures exist in
the the B meson wavefunctions. $\phi_{B}(\mathbf{k})$ and
$\bar{\phi}_{B}(\mathbf{k})$ satisfy the following normalization
conditions respectively
\begin{eqnarray}
\int\frac{d^{4}k}{(2\pi)^4}\phi_{B}(\mathbf{k})=\frac{f_{B}}{2\sqrt{2N_{c}}},\;\;\;
\int\frac{d^{4}k}{(2\pi)^4}\bar{\phi}_{B}(\mathbf{k})=0.
\end{eqnarray}
In the numerical calculation, one usually ignores the contribution
of $\bar{\phi}_{B}(\mathbf{k})$ \cite{ignore-b,Cai-dian} and only
take the contribution from
\begin{eqnarray}
\Phi_{B}&=&\frac{1}{\sqrt{2N_{c}}}(\not p
+m_{B})\gamma_{5}\phi_{B}(\mathbf{k}).
\end{eqnarray}
The wave function of B meson is given as \cite{ignore-b,Cai-dian}
\begin{eqnarray}
\phi_{B}(x,b)&=&\frac{N_{B}}{2\sqrt{2N_{c}}}f_{B}x^2 (1-x)^2
\exp\Big[-\frac{M_{B}^{2}x^{2}}{2\omega_{b}^{2}}-\frac{1}{2}(b
\omega_{b})^{2}\Big],
\end{eqnarray}
where $\omega_{b}=0.4$ GeV and $N_{B}=2.4\times 10^{3}$. The decay
constant of B meson $f_{B}=0.18$ GeV.

The twist-3 light cone distribution amplitude of K meson is
expressed as %(Phys. Rev. D 72, 114005 (2005))
\begin{eqnarray}
\langle \bar{K}^{0}(p')|\bar{s}_{\alpha}(z)q_{\beta}(0)|0\rangle
=\frac{i}{\sqrt{2N_{c}}}\int^{1}_{0}dxe^{ix{p'}\cdot
{z}}\big\{\gamma_{5}\not p'
\phi^{A}_{K}(x)+m_{0}^{K}\gamma_{5}\phi^{P}_{K}(x)+m_{0}^{K}[\gamma_{5}(\not
v \not n -1)]\phi_{K}^{T}(x)\big\}_{\beta\alpha}\label{k meson},
\end{eqnarray}
where $m^{K}_{0}=\frac{m^{2}_{K}}{m_{s}+m_{d}}$. In a recent work
\cite{Ball}, the K meson wave function distribution amplitudes
used in eq. (\ref{k meson}) is given as
\begin{equation}
\phi_{K}^{A}(x) = {f_K \over 2 \sqrt{2 N_c}} \{6 x (1-x) \left( 1
+
  a^K_{1}   C_{1}^{3/2}(2x-1)+  a^K_{2}
  C_{2}^{3/2}(2x-1)\right)\},
\end{equation}

\begin{eqnarray}
\lefteqn{\phi_{K}^{P}(x) = {f_K \over 2 \sqrt{2 N_c}}\{1 +
3\rho^K_+ (1+6 a^K_2) - 9 \rho^K_- a^K_1 + C_1^{1/2}(2x-1)
\left[\frac{27}{2}\,\rho^K_+ a^K_1 - \rho^K_- \left( \frac{3}{2} +
27 a^K_2\right)\right]}
\nonumber\\
&& + C_2^{1/2}(2x-1) \left( 30 \eta_{3K} + 15 \rho^K_+ a^K_2 - 3
\rho^K_- a^K_1\right) + C_3^{1/2}(2x-1)\left( 10 \eta_{3K}
\lambda_{3K} - \frac{9}{2}\,\rho^K_- a^K_2\right)
\nonumber\\
&&{} - 3\eta_{3K} \omega_{3K} C_4^{1/2}(2x-1) +
\frac{3}{2}\,(\rho^K_+ + \rho^K_-) (1-3a^K_1+6a^K_2)\ln x
\nonumber\\
&&{} + \frac{3}{2}\,(\rho^K_+-\rho^K_-)(1+3 a^K_1 + 6  a^K_2) \ln \bar x \},\\
\lefteqn{\phi^{T}_{K}=\frac{1}{6}\frac{d \phi_{K}^{\sigma}(x)}{d
x}\,,}\\
 \lefteqn{\phi_{K}^{\sigma}(x) = {f_K \over 2 \sqrt{2 N_c}}\{6x\bar x \left[ 1 +
\frac{3}{2}\,\rho^K_+ + 15 \rho^K_+ a^K_2 - \frac{15}{2}\,\rho^K_-
a^K_1 + \left( 3 \rho^K_+ a^K_1 - \frac{15}{2}\,\rho^K_-
a^K_2\right) C_1^{3/2}(2x-1)\right.}
\nonumber\\
&&{}\left. +
    \left(5\eta_{3K} -\frac{1}{2}\,\eta_{3K}\omega_{3K}  +
    \frac{3}{2}\,\rho^K_+ a^K_2 \right)
    C_2^{3/2}(2x-1) + \eta_{3K} \lambda_{3K}
    C_3^{3/2}(2x-1)\right]
\nonumber\\
&&{} +  9 x \bar x (\rho^K_++\rho^K_-) (1-3 a^K_1+6 a^K_2) \ln x +
9 x \bar x (\rho^K_+-\rho^K_-) (1+3 a^K_1+6 a^K_2) \ln \bar x \}
\,,
\end{eqnarray}
with
\begin{eqnarray}
    \rho_{+}^K &=&\frac{(m_s+m_q)^2}{m^2_K},\qquad \rho_{-}^K =
\frac{m_s^2-m_q^2}{m_K^2}, \qquad \eta_{3K}=
\frac{f_{3K}}{f_K}\,\frac{m_q+m_s}{m_K^2},\nonumber\\
C^{1/2}_{1}(t)&=&t ,\qquad
C^{1/2}_{2}(t)=\frac{1}{2}(3t^{2}-1),\qquad
C^{1/2}_{3}(t)=\frac{1}{2}({5t^{3}}-{t}),\qquad
C^{1/2}_{4}(t)=\frac{1}{8}(3-30t^{2}+35t^{4}),\nonumber\\
C^{3/2}_{1}(t)&=&3t,\qquad
C_{2}^{3/2}(t)=\frac{3}{2}(5t^{2}-1),\qquad
C_{3}^{3/2}=\frac{1}{2}({35 t^3}+{3t}),\nonumber
\end{eqnarray} where $a_1^K=0.06\pm 0.03$, $a_2^K =
0.25\pm 0.15$, $f_{3K}= (0.45\pm 0.15)\times 10^{-2} $ GeV$^{2}$,
$\omega_{3K}= -1.2\pm 0.7$, $\lambda_{3K}= 1.6\pm 0.4$,
$f_{K}=0.16$ GeV, $m_{s}=137\pm 27$ MeV, $m_{q}=5.6\pm 1.6$ MeV.

The light cone distribution amplitude of $h_{c}$ meson is proposed
in \cite{K.C. Yang}
\begin{eqnarray}
\langle
h_{c}(q,\epsilon)|\bar{c}_{\alpha}(z)c_{\beta}(0)|0\rangle=-\frac{i}{\sqrt{2N_{c}}}
\int^{1}_{0} du e^{iu {q}\cdot {z}}\bigg\{
f_{h_{c}}m_{h_{c}}{\Big[\not\epsilon^{*}_{\parallel}+\frac{m^{2}_{h_{c}}\not
z \epsilon^{*}\cdot z}{2({q}\cdot
{z})^2}\Big]}\gamma_{5}\phi_{h_{c}}^{\parallel}(u)-f^{\perp}_{h_{c}}\not
\epsilon^{*}_{\perp}\not q
\gamma_{5}\phi_{h_{c}}^{\perp}(u)\bigg\}_{\beta\alpha}\label{hc-wave}
\end{eqnarray}
with
$$\epsilon_{\parallel\mu}^{*}=\frac{\epsilon^{*}\cdot z}{q\cdot z}
\Big[q_{\mu}-\frac{m^{2}_{h_{c}}z_{\mu}}{q\cdot
z}\Big],\qquad\epsilon^{*}_{\perp\mu}=\epsilon^{*}_{\mu}
-\epsilon^{*}_{\parallel\mu}.$$ We have dropped out the twist-3
distribution amplitudes here, because  $h_{c}$ is heavy, the higher
twist contributions are negligible. Because the produced $h_{c}$ in
the transition  $B^{-}\to h_{c}K^{-}$ can only be longitudinally
polarized, the wavefunction $\phi_{h_{c}}^{\perp}(u)$ does not
contribute and we neglect its explicit form in the text. It is a
good approximation to assume that $\phi_{\parallel}(u)$ is of the
same form as $\chi_0^v$ {which is the leading-twist distribution
amplitude of $\chi_{c0}$ defined in }Eq.(59) of Ref.
\cite{hc-amplitude}
\begin{eqnarray}
\phi_{h_c}^{\parallel}(x)&=&{27.46\frac{f_{h_{c}}}{2\sqrt{2N_{c}}}(1-2x)\bigg\{
\frac{x(1-x)[1-4x(1-x)]}{[1-2.8x(1-x)]^2 }\bigg\}^{0.7}},
%\nonumber\\
%\phi_{\perp}(u)&=&27.46\frac{f_{h_{c}}}{2\sqrt{2N_{c}}}u(1-u)\bigg\{
%\frac{x(1-x)[1-4x(1-x)]}{1-2.8x(1-x)^2 }\bigg\}^{0.7}
\end{eqnarray}
where $h_{c}$ decay constant $f_{h_{c}}$ is set to be 0.335 GeV
which is the same as $f_{\chi_{c1}}$ as has been done in Ref.
\cite{suzuki}.

\end{document}